\documentclass[a4paper,nodate]{sab} 

\usepackage{graphicx} 
\usepackage{txfonts,natbib} 
\bibpunct{(}{)}{;}{a}{}{,} 
\usepackage{hyperref}
\usepackage{url}
\usepackage[T1]{fontenc} 
\usepackage{t1enc} 
\renewcommand\ackname{Acknowledgements.} 
\renewcommand\andname{\&}
\renewcommand\keywordname{Keywords.} 
\renewcommand\figureptname{Figure}
\renewcommand\tableptname{Table}

\renewcommand\noteaddname{Note added to proofs.}
\renewcommand\andname{\&}

\idline{Boletim da Sociedade Astron\^omica Brasileira, {\bf 31}, no. 1, XX-XX}{1}
\begin{document} 
\title{Photometric candidate selection and spectroscopic confirmation of new PNe and SySts in the Galactic plane} 

\author{Giovanna Liberato \inst{1}, Denise R. Gonçalves \inst{1}, Luis A. Gutiérrez-Soto \inst{2} \and Stavros Akras \inst{3}} 
\institute{Observatório do Valongo - Universidade Federal do Rio de Janeiro. \email{giovanna19@ov.ufrj.br}   \and Instituto de Astronomia, Geofísica e Ciências Atmosféricas - Universidade de São Paulo  \and Institute for Astronomy, Astrophysics, Space Applications and Remote Sensing, National Observatory of Athens.}
\date{Received } 

\Abstract {About 3,500 planetary nebulae (PNe) are currently known in the Milky Way, which shows a great discrepancy with the expected number for these objects, regardless of the reference used, $33-59 \times 10^{3}$.  The same holds for symbiotic stars (SySts) as well, since the expected number in the Galaxy ($3-400 \times 10^{3}$) differs considerably from the amount of known ones (approximately 300).
Studies on PNe and SySts are of great importance because they provide vital clues to the understanding of the late-stage of stellar evolution for low-to-intermediate mass stars. In addition, these classes of objects play a large role in the chemical evolution of the Galaxy through the ejection
of their material to the interstellar medium (ISM), enriching it with the various chemical elements produced throughout their evolution. This project aims to contribute to the detection of new PNe and SySts in
the Galaxy, thus decreasing the discrepancy between the observed and theoretical populations.
Using simultaneously the third data release from the optical survey VPHAS+ (The VST Photometric H$\alpha$ Survey of the Southern Galactic Plane and Bulge), which maps the southern hemisphere of the
Galaxy's plane with the $r$, $i$, and H{$\alpha$} filters, and the IR colors of the catalog AllWISE (Wide-field Infrared Survey Explorer
+ The Two Micron All Sky Survey), we end up with a number of PN and SySt candidates. Subsequently, we confirm the nature of these objects through spectroscopic observations at the SOAR telescope (Southern Astrophysical Research Telescope). So far, we have selected PN candidates and performed the spectroscopic follow-up of 8 of them. In this presentation we show the project's preliminary results, which consist of the discovery of at least one new planetary nebula, and other emission-line sources still to be confirmed either as PN or SySt.}{Aproximadamente 3.500 nebulosas planetárias (PNe) são conhecidas na Via-Láctea atualmente, o que mostra uma grande discrepância com o número esperado para esses objetos, independentemente da referência utilizada, $ 33 - 59 \times 10^{3}$ . Essa situação também se faz presente no caso das estrelas simbióticas (SySts), uma vez que o número esperado na Galáxia ($3 - 400 \times 10^{3}$) difere consideravelmente do número de conhecidas (aproximadamente 300).
Estudos acerca de PNe e SySts são de grande importância, porque fornecem pistas vitais para a compreensão da evolução estelar em estágio avançado das estrelas de massas baixa e intermediária. Além disso, essas classes de objetos desempenham um papel crucial na evolução química da Galáxia através da ejeção de seu material para o meio interestelar (ISM), enriquecendo-o com os diversos elementos químicos produzidos ao longo de sua evolução. Este projeto objetiva contribuir para a detecção de novas PNe e SySts na Galáxia, e, portanto, diminuir a discrepância entre as populações observadas e teóricas.
Usando, simultaneamente, o terceiro lançamento de dados do levantamento óptico VPHAS+ (The VST Photometric H$\alpha$ Survey of the Southern Galactic Plane and Bulge), o qual mapeia o hemisfério sul do plano Galáctico com os filtros $r$, $i$ e H$\alpha$,  e as cores do catálogo infravermelho AllWISE (Wide-field Infrared Survey Explorer + The Two Micron All Sky Survey), selecionamos candidatos a PN e SySt. Posteriormente, confirmamos a natureza desses através de observações espectroscópicas no telescópio SOAR (Southern Astrophysical Research Telescope).
Até o momento, selecionamos os candidatos a PN e realizamos as observações espectroscópicas de 8 deles. Nesta apresentação mostramos os resultados preliminares deste projeto, que consistem na descoberta de pelo menos uma nova nebulosa planetária, e outras fontes de linhas de emissão que ainda serão confirmadas como PN ou SySt.}



\authorrunning{G. Liberato et al.} 
\titlerunning{New PN and SySt candidates in the Galactic plane}

\maketitle 

\section{Introduction}

PNe are the fossils of low- and intermediate-mass stars, with masses between about 0.8 and 8.0  $\mathrm{M_\odot}$. After consuming most of their hydrogen, the stars evolve to the red giant and subsequently to the AGB phases producing slow winds  (30 km/s, Ramstedt et al., 2020). After exposing their hot core, they enter the post-AGB phase increasing their temperature and generating a tenuous but fast stellar wind (Guerrero \& De Marco, 2013). The slow winds push and compress the AGB material into a dense shell that expands towards the surrounding medium. At the same time, the central star produces UV photons that ionize the shell. Due to this expansion, the density of PNe decreases with time, resulting in gradually declined emission. As a consequence of this process PNe quickly become undetectable. The PN phase is very short (20,000 - 30,000 years) compared to the entire lifetime of a low-mass or
intermediate-mass star (typically $10^{9}$ years). At the end of the evolution of these stars, their outer layers are ejected resulting in the enrichment of the ISM with heavy elements (Karakas \& Lattanzio,
2014). The ionized gas contains information about the chemical composition of when and where the progenitor stars were formed and also provides clues to chemical evolution via stellar nucleosynthesis.

SySts are interacting binary systems that exhibit a composite spectrum with emission lines from H I and He I and often [O III] and He II together with absorption signatures from the evolved giant stars (Kenyon, 1986; Belczyński et al., 2000). The typical configuration of a SySt consists of a red giant transferring material to a white dwarf star (WD) through the stellar wind. Part of this wind is ionized by the WD radiation, producing a spectrum with two components: one of them described by the absorption characteristics of the giant's cold stellar photosphere and the other by the emission lines of excited ions (e.g., Corradi et al., 2008). 

SySts are important objects in Astrophysics because they allow the study of the physical mechanism leading to: supersoft X-ray sources (Jordan et al., 1996); thermonuclear explosions (Munari, 1997);  collimation of stellar winds; and formation of jets (Tomov,
2003) and also the jets' relation with the formation of bipolar PNe (Corradi et al., 2003); among others. Another important aspect of SySts is that they are probably progenitors of Type~Ia supernovae (Whelan \& Iben, 1973; Chen et al., 2011).

In the Galaxy, there is a significant discrepancy between the expected number of PNe, $33-59 \times 10^{3}$ (Moe \& De Marco, 2006), and the observed, 3,500 (Viironen et al., 2009a,b; Parker et al. 2016; Fragkou et al., 2018) and between the expected number of SySts ($3-400 \times 10^{3}$) 
versus the known objects (approximately 300, Akras et al., 2019a; Merc et al., 2019 online catalog). 

The VPHAS+ survey provides photometric measurements in the $r$, $i$ and H$\alpha$ bands, making it ideal for searching for H$\alpha$ emitters, such as PNe and SySts. Machine learning techniques are essential tools to deal with the large amount of data in the catalogs and search for possible new objects of specific groups, such as PNe and SySts. Akras et al. (2019b,c) explored the application of two machine learning techniques for the identification of new SySts and PNe candidates. In particular, the decision tree and the nearest neighbors algorithm (KNN, k nearest neighbors) were applied in a 10-dimensional space of color indices of the infrared data from the AllWISE catalog in order to find new criteria that discriminate PNe and SySts from 
objects that resemble their colours, in an efficient way.

\section{PN candidate selection}

To select PN candidates, we use the VPHAS+ DR3 and the AllWISE catalogues. The first step is to restrict our candidates to objects: i) with magnitude in the $r$-band $<$~19.5; ii) with errors $<$ 0.1, for the optical magnitudes, as proposed by the VPHAS+ team; and iii) errors $<$ 0.3 for the infrared magnitudes, to select only objects whose magnitudes are well measured. We then use the VPHAS+ $r$, $i$ and H$\alpha$ magnitudes to, based on the positions of known PNe in color-color diagrams proposed by Viironen et al. (2009a,b), identify mild and strong H$\alpha$ emitters like PNe. The H$\alpha$ emitters definition is as follows.

 \begin{itemize}
    \setlength\itemsep{0.01em}
     \item Optical VPHAS+ Zone 1 \\
{\it r}-H$\alpha$ $>$ 0.25({\it r-i}) + 1.9
   \end{itemize}

 \begin{itemize}
    \setlength\itemsep{0.01em}
     \item Optical VPHAS+ Zone 2 \\
 0.25({\it r-i}) + 0.87) $<$ {\it r}-H$\alpha$ $<$ 0.25({\it r-i}) + 1.9
   \end{itemize}

\begin{figure}
    \centering
    \includegraphics[width=0.95\linewidth]{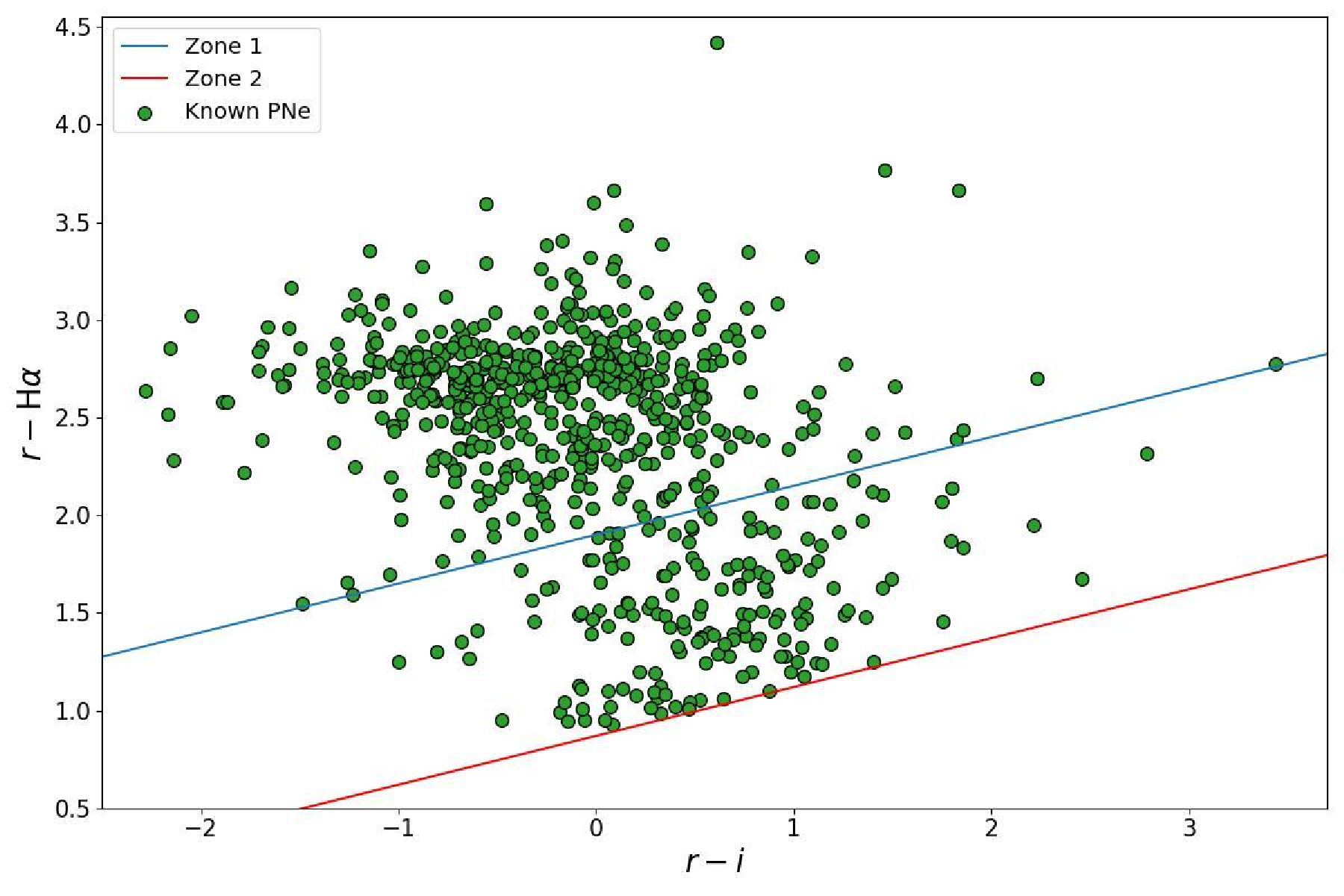}
    \caption{Color-color diagram for known PNe selected through the optical criteria. Blue and red lines delimit zones 1 and 2, respectively.}
    \label{fig:pne}
\end{figure}

Applying these color criteria we selected the objects present in Zone 1 and Zone 2 (as defined by Viironen at al., 2009a,b). The main difference between the two zones in that candidates in Zone 1 have a higher probability to be a genuine, compact PNe (Fig.~\ref{fig:pne}).

In addition, we use the AllWISE/2MASS W1, W2, W3, W4, J, H and K magnitudes of the objects selected in VPHAS+ to apply the infrared color criteria proposed by Akras, Guzman-Ramirez $\&$ Gonçalves (2019c) via the classification tree (machine learning) method (see their Fig.~1). Thus, the IR criteria used in this work are as below.

 \begin{itemize}
    \setlength\itemsep{0.01em}
     \item Model 1 \\
     W1-W4$\geq$7.87 \& J-H<1.10
   \end{itemize}

    \begin{itemize}
    \setlength\itemsep{0.01em}
     \item Model 2 \\
     H-W2$\geq$2.24 \& J-H<0.503
   \end{itemize}

    \begin{itemize}
    \setlength\itemsep{0.01em}
     \item Model 3 \\
     J-H<1.31 \& K-W3$\geq$6.42
   \end{itemize}

Once the PN candidate list is constructed from the above color criteria, we performed a cross-match between our objects and the HASH PN catalogue (HASH Hong Kong/AAO/Strasbourg H$\alpha$ planetary nebula database) to highlight which objects are already identified as PNe. 

It is also necessary to carry out a visual inspection of the candidates, eliminating those that do not look like good PN candidates, and could, possibly, represent mimics.  The most common PN and SySt mimics are compact H~II regions, Herbig-Hero objects, young stellar objects, cataclysmic variables, Wolf-Rayet stars, supernova remnant, T-Tauri stars, among others (also see Gutiérrez-Soto, 2019). The  visual inspection was done with the Aladin Sky Atlas. 

\section{SySt candidate selection}

The procedure of selecting  SySt candidates is very similar to the PN candidates selection process. First, we apply the optical color criterion (Corradi et al., 2008) developed to select $\mathrm{H}\alpha$ emitters:

   \begin{itemize}
    \setlength\itemsep{0.01em}
     \item Optical VPHAS+ \\ ({\it r-H$\alpha$})$\geq$0.25$\times$({\it r-i} ) + 0.65 ~.
   \end{itemize}

Subsequently, taking the results from the above procedure we apply the IR color criteria obtained by Akras et al. (2019b), which are based on classification tree techniques using as training sample the groups of all known Galactic SySts and their mimics (see Fig. 7 of Akras et al., 2019b), as below: 

   \begin{itemize}
    \setlength\itemsep{0.01em}
     \item IR criterion for SySts \\
     J-H$\geq$0.78 \& 0<Ks-W3<1.18 \& W1-W2<0.09 \\ or \\
     J-H$\geq$0.78 \& 0<Ks-W3<1.18 \& W1-W2$\geq$0.09 \\ 
     \& 0<W1-W4<0.92 ~.
   \end{itemize}

In addition, we use the IR criterion which allows separating SySt candidates from single giant stars of type K and M (Akras et al. 2019b, see their Fig. A5).

\begin{itemize}
    \setlength\itemsep{0.01em}
     \item IR criterion among SySts K- and M-giants \\
     H-W2$\geq$0.206 \& Ks-W3$\geq$0.27 
\end{itemize}
   
It is common that these stars contaminate the list of candidates. The reason is that K and M stars are the cold companions in the symbiotic binary systems. As in the case of the PN candidates, the visual inspection is the last step of the SySt candidate selection.

\section{Preliminary results}

\begin{figure}[!htb]
    \centering
    \includegraphics[width=0.9\linewidth]{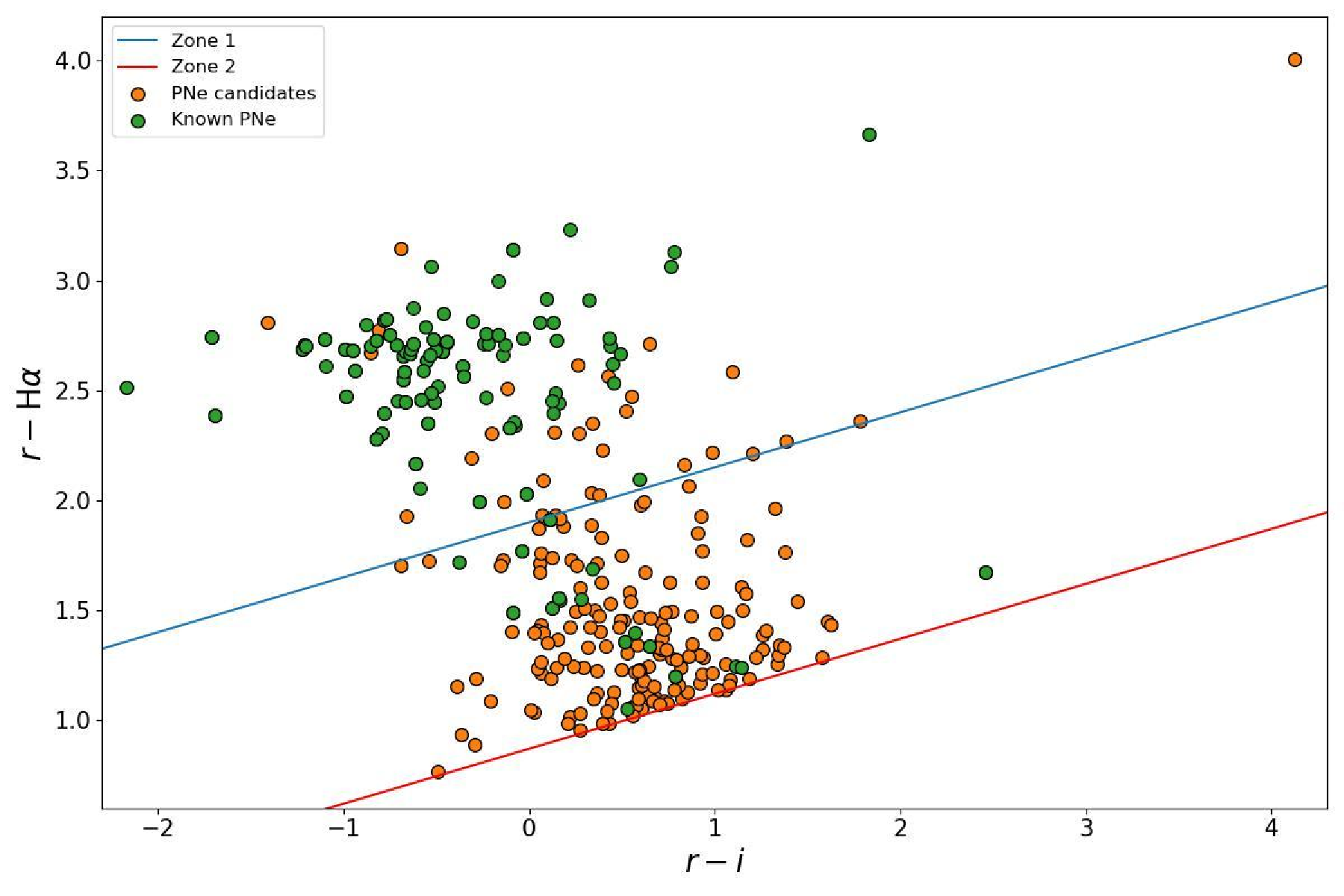}
    \caption{Color-color diagram of objects that pass through the color criteria applied. The points in green are already known PNe. Blue and red lines delimit zones 1 and 2, respectively.}
    \label{fig:pneeu}
\end{figure}

By adopting the 5.7~arcsec aperture for VPHAS+ and the PSF one for the ALLWISE photometry, the procedure described above returned 344 PN candidates. These objects were, then, correlated with the catalog of known planetary nebulae, HASH catalogue\footnote{The University of Hong Kong/Australian Astronomical Observatory/Strasbourg Observatory $H\alpha$ Planetary Nebula}, revealing that 159 of them are already known PNe, as it can be seen in Fig.~\ref{fig:pneeu}.

\begin{figure}
    \centering
    \includegraphics[width=0.755\linewidth]{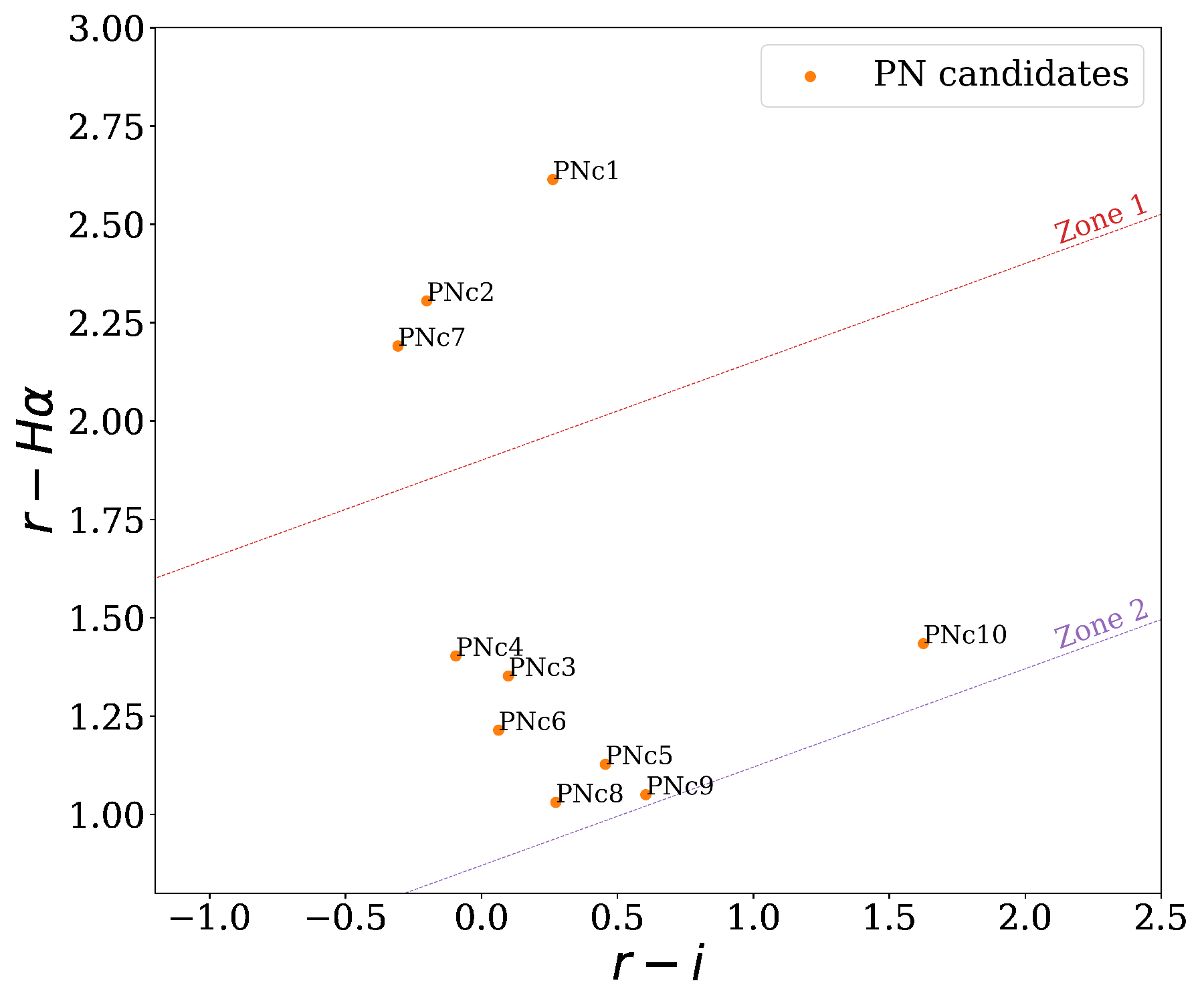}
    \includegraphics[width=0.755\linewidth]{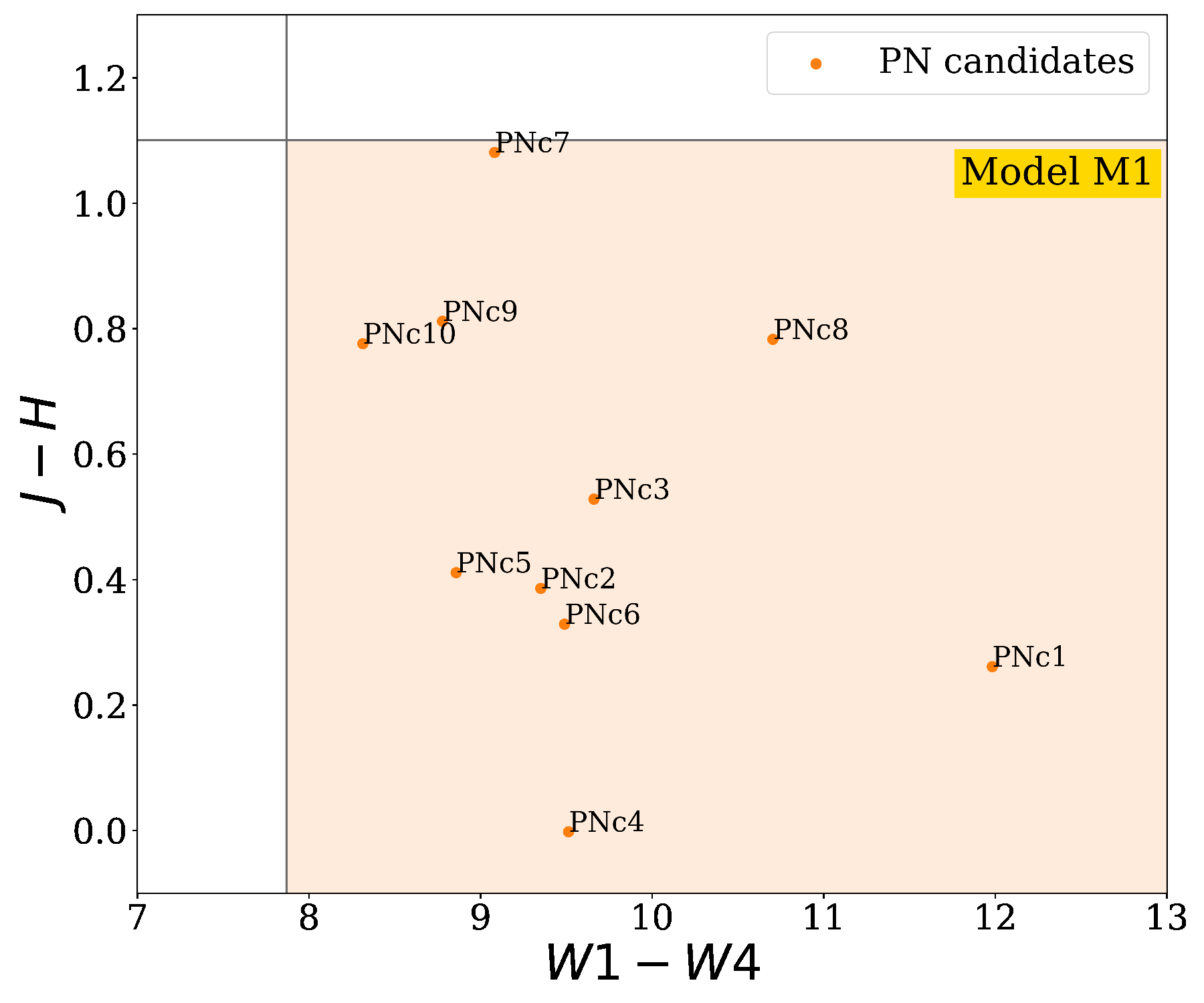}
    \includegraphics[width=0.755\linewidth]{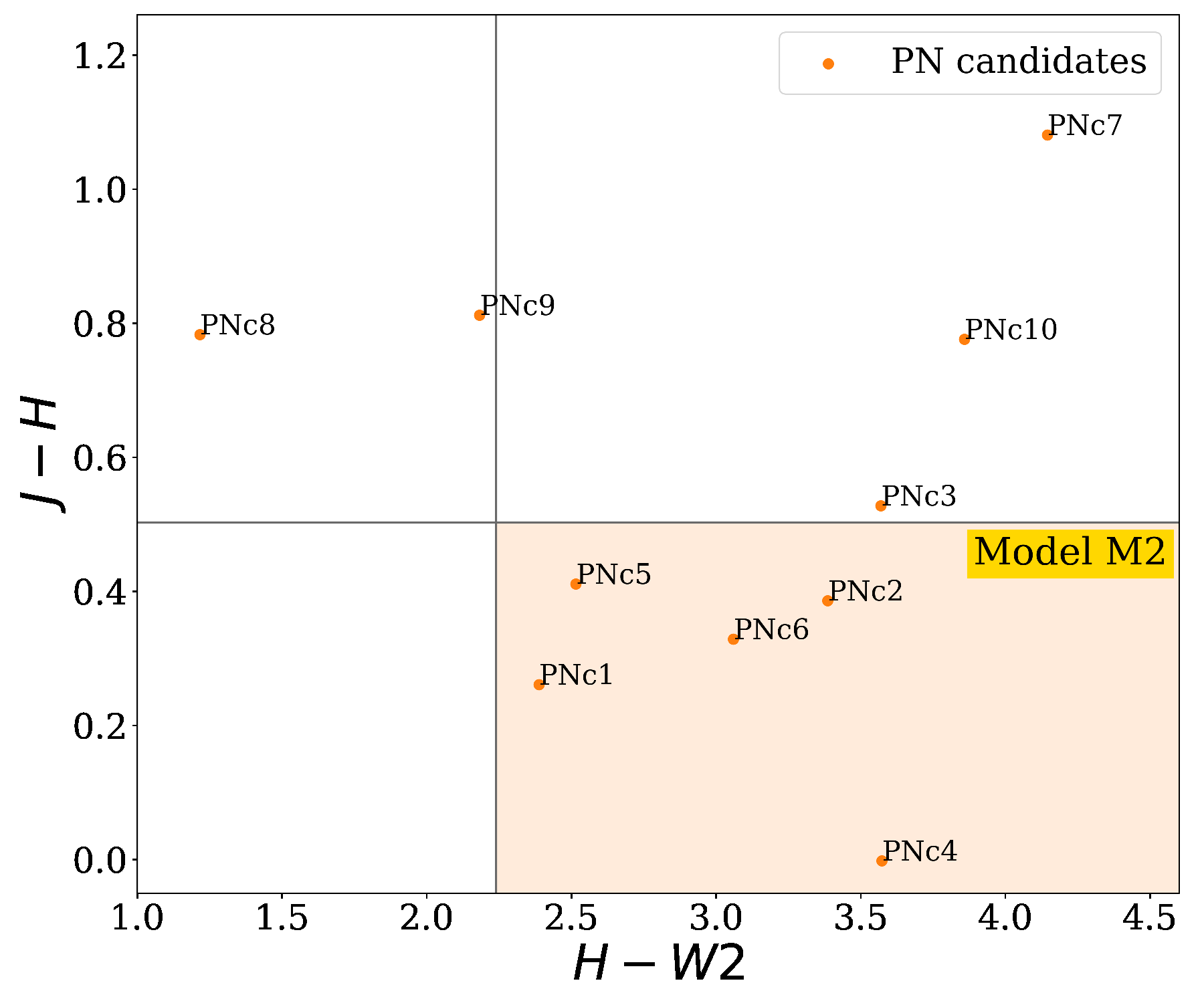}
    \includegraphics[width=0.755\linewidth]{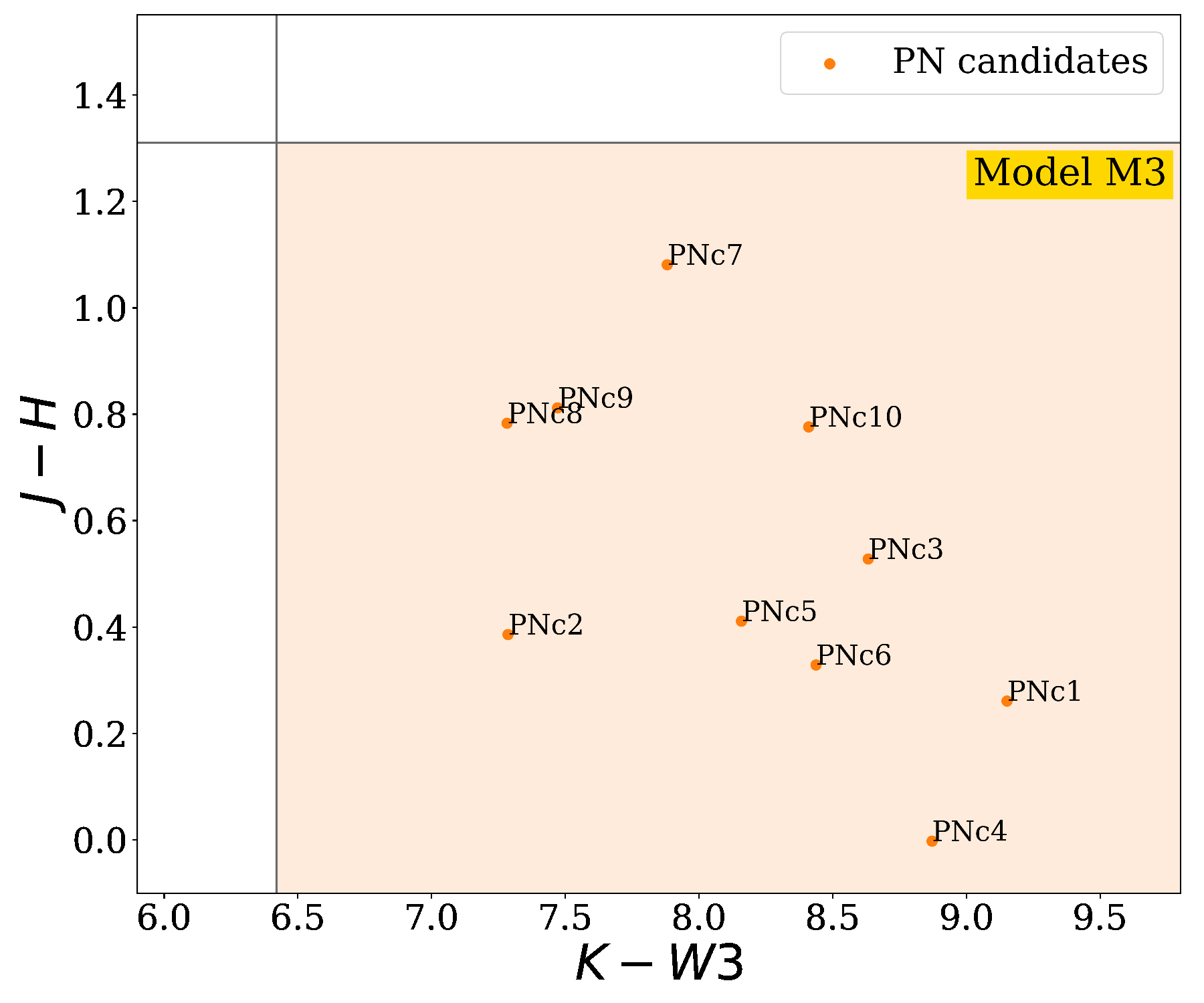}
    \caption{Optical (topmost) and IR color-color diagrams (top-second, middle and bottom), referring to models M1, M2 and M3, respectively, of the ten PN candidates chosen for spectroscopic follow-up.}
    \label{fig:proposta}
\end{figure}

\begin{figure}
    \centering
    \includegraphics[width=0.8\linewidth]{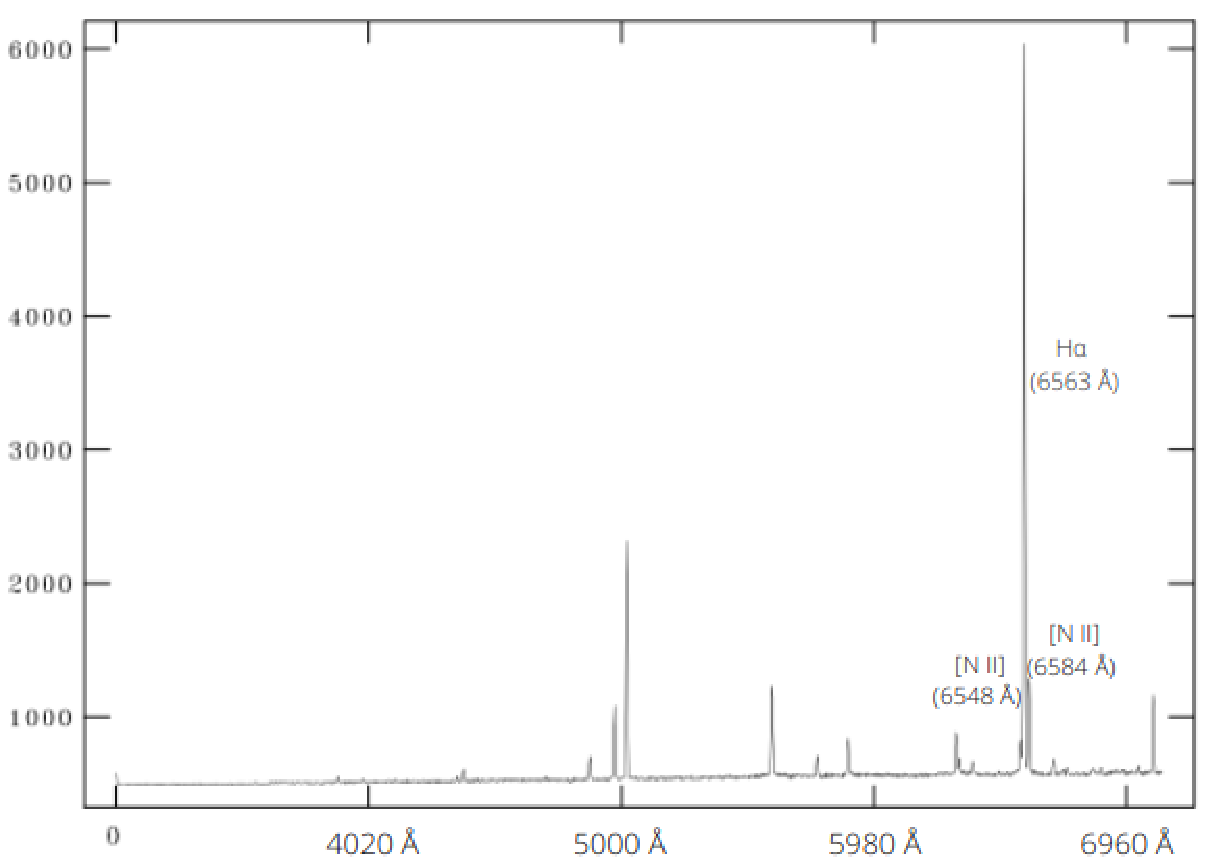}
    \includegraphics[width=0.8\linewidth]{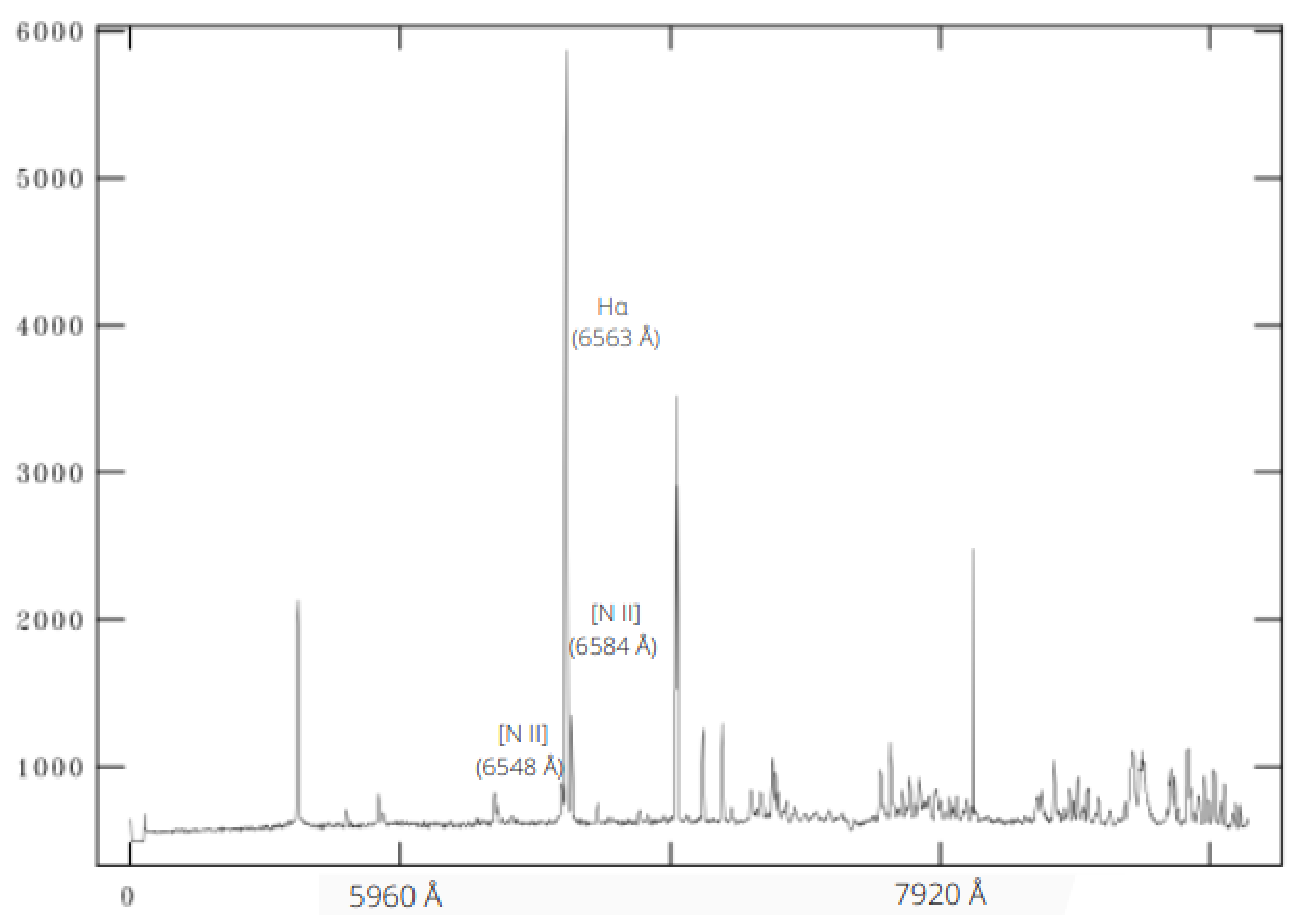}
    \includegraphics[width=0.8\linewidth]{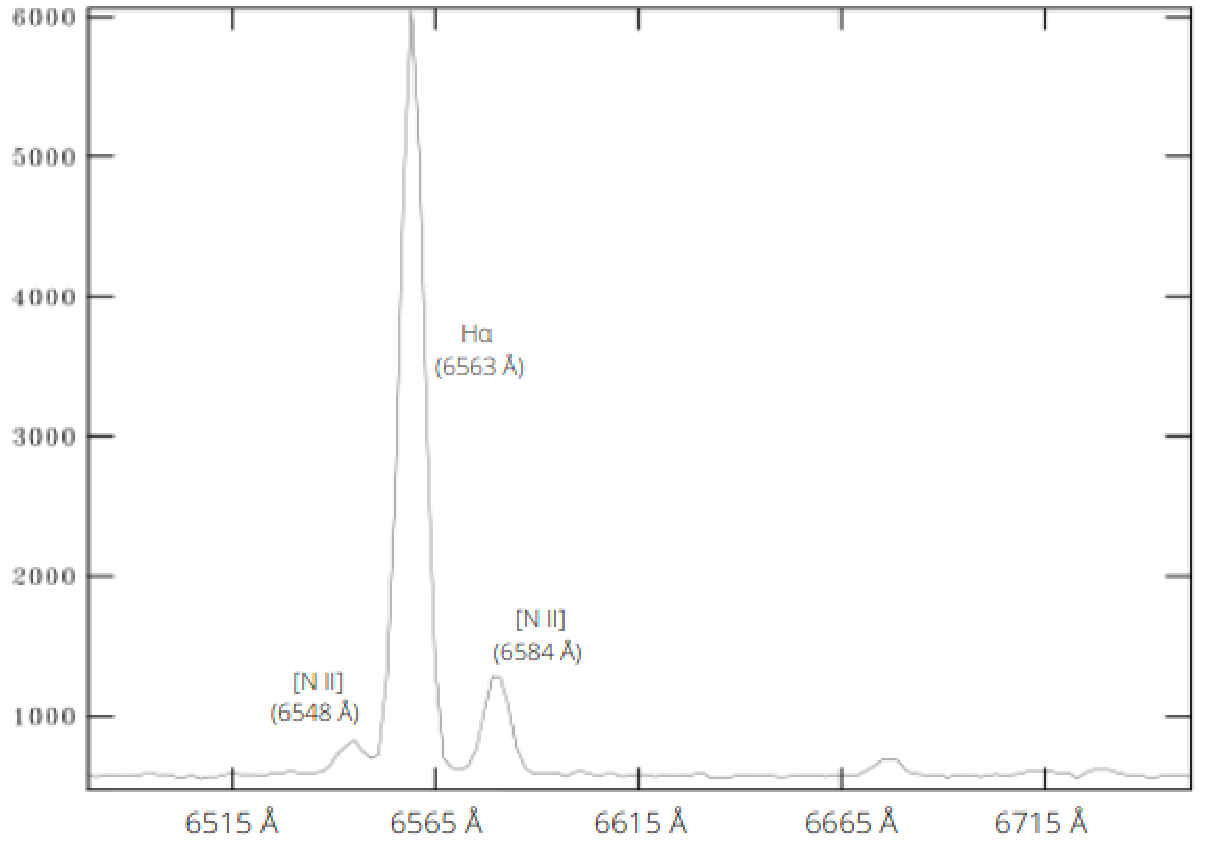}
    \caption{Goodman@SOAR longslit spectrum of a new PN discovered in the present work. The spectrum was obtained with a 1.2~arcsec slit width and the 400 lines/mm grating, providing a dispersion of 1 Å/pix. The wavelength range 3000-9050Å was covered using two gratings, M1:3000-7050$\AA$ and M2:5000-9050$\AA$. From top to bottom, the blue (M1), red (M2) and the $H\alpha$-[N~{\sc ii}] portions of the PN spectrum.}
    \label{fig:pnc07}
\end{figure}

\begin{figure}
    \centering
    \includegraphics[width=0.8\linewidth]{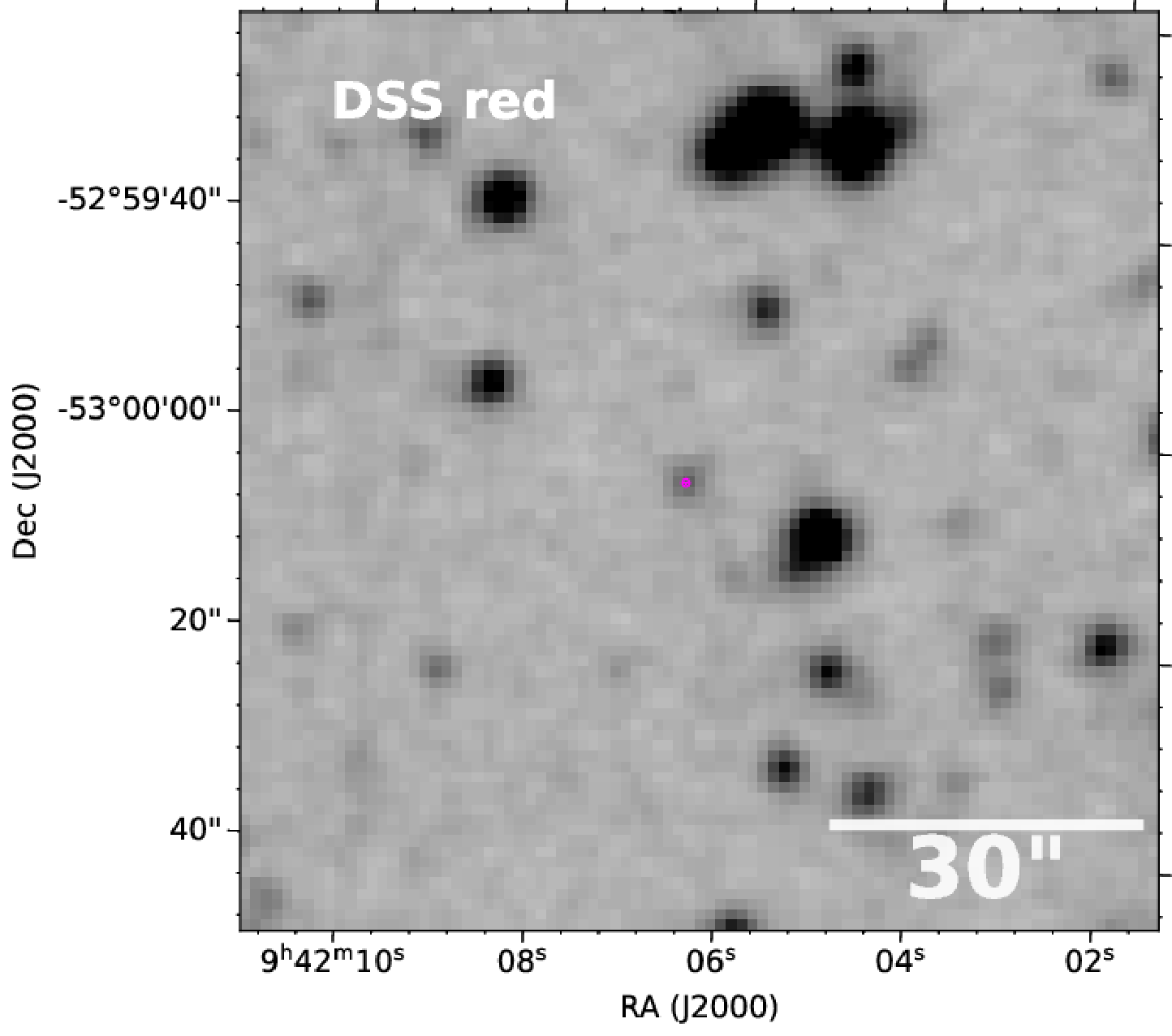}
    \includegraphics[width=0.8\linewidth]{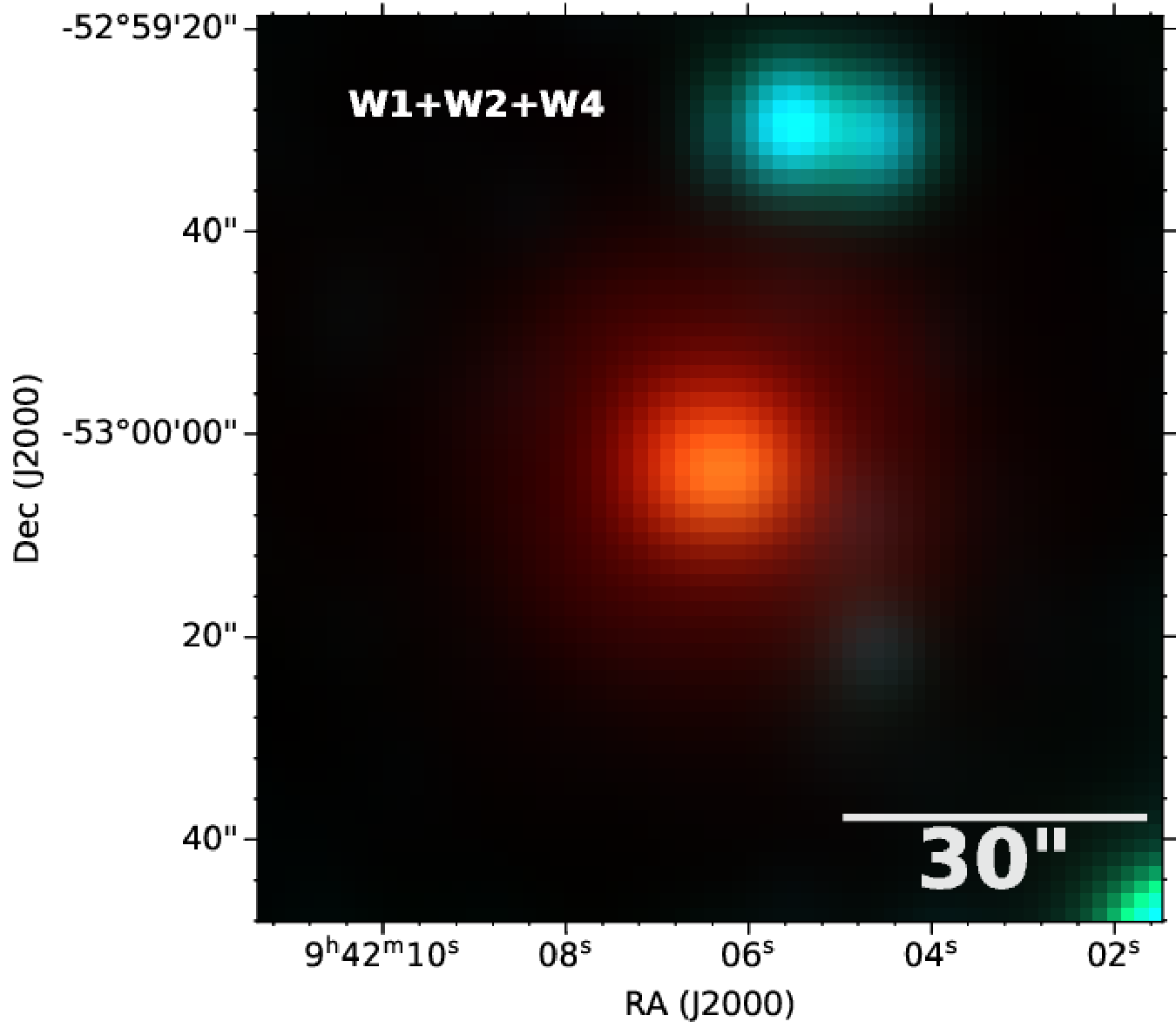}
    \caption{Images of the first PN discovered in the present project, on optical (DSS red) and infrared  light (composition of the W1, W2 and W4 ALLWISE detections).}
    \label{fig:pn7}
\end{figure}

In order to confirm the nature of the 185 sources, we selected the best 10 PN candidates, based on their observability, position in the color-color diagrams and visual inspection, for optical spectroscopic follow-up observations, see Fig.~\ref{fig:proposta}. The observations were carried out with the SOAR telescope, by using the Goodman High Throughput Spectrograph, in the longslit mode, on January 2022. Eight of the PN candidates were observed, and the analysis is still ongoing. Final numbers of discoveries are not available yet, since we did not finish the calibrations and analyses. Even though, as the spectra in Fig.~\ref{fig:pnc07} shows -- the blue and red part of the optical spectrum of a candidate --, we have identified a few emission-line evolved stars plus their nebulosity, either PNe or SySts. The source in Fig.~\ref{fig:pnc07} and Fig.~\ref{fig:pn7} is a new genuine planetary nebula, the first discovery in the present project.

Currently, we are working on the selection of the SySt candidates. We are testing magnitude apertures in order to decide which one returns the better candidate list. New observational follow-ups will be performed for a reasonable number of both PN and SySt candidates in the near future. 

\begin{acknowledgements}Authors acknowledge the following financial supports: GLC, FAPERJ fellowship (203.691/2021); DGR,  CNPq grants 428330/2018-5 and 313016/2020-8; SA,  the grant 5077 financed by IAASARS/NOA.
Based on data products from observations made with ESO Telescopes at the La Silla Paranal Observatory under programme ID 177.D-3023, as part of the VST Photometric $H\alpha$ Survey of the Southern Galactic Plane and Bulge (VPHAS+, www.vphas.eu). 
This publication makes use of data products from the Wide-field Infrared Survey Explorer, which is a joint project of the University of California, Los Angeles, and the Jet Propulsion Laboratory/California Institute of Technology, and NEOWISE, which is a project of the Jet Propulsion Laboratory/California Institute of Technology. WISE and NEOWISE are funded by the National Aeronautics and Space Administration
Based on observations obtained at the Southern Astrophysical Research (SOAR) telescope, which is a joint project of the Ministério da Ciência, Tecnologia e Inovações do Brasil (MCTI/LNA), the US National Science Foundation’s NOIRLab, the University of North Carolina at Chapel Hill (UNC), and Michigan State University (MSU).
\end{acknowledgements} 

\newpage

\end{document}